\documentclass{article}

\usepackage{arxiv}

\usepackage[utf8]{inputenc} 
\usepackage[T1]{fontenc}    
\usepackage{hyperref}       
\usepackage{url}            
\usepackage{pdflscape}
\usepackage{longtable}
\usepackage{booktabs}       
\usepackage{array}
\usepackage{amsfonts}       
\usepackage{nicefrac}       
\usepackage{microtype}      
\usepackage{amsmath}
\usepackage{cleveref}       
\usepackage{lipsum}         
\usepackage{graphicx}
\usepackage[square,sort,comma,numbers]{natbib}
\usepackage{doi}
\usepackage{float}
\usepackage{tabularx}
\usepackage{subcaption}
\usepackage{caption} 

\newcolumntype{C}[1]{>{\centering\arraybackslash}p{#1}} 

\title{Machine Learning-Based Detection of DDoS Attacks in VANETs for Emergency Vehicle Communication}

\newif\ifuniqueAffiliation
\uniqueAffiliationtrue

\ifuniqueAffiliation 
\author{ \href{https://orcid.org/0009-0002-7235-8653}{\includegraphics[scale=0.06]{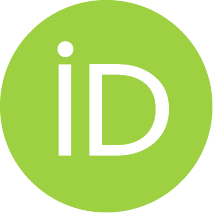}\hspace{1mm}Bappa~Muktar}\thanks{Use footnote for providing further
		information about author (webpage, alternative
		address)---\emph{not} for acknowledging funding agencies.} \\
	Department of Computer Science\\
	University of Quebec in Outaouais (UQO)\\
	Gatineau, QC J8X 3X7 \\
	\texttt{mukb06@uqo.ca} \\
	\And
	\href{https://orcid.org/0009-0009-0944-9721}{\includegraphics[scale=0.06]{orcid.pdf}\hspace{1mm}		Vincent.~Fono} \\
	Department of Computer Science\\
	University of Quebec in Outaouais (UQO)\\
	Gatineau, QC J8X 3X7 \\
	\And
	\href{https://orcid.org/0000-0003-0324-1443}{\includegraphics[scale=0.06]{orcid.pdf}\hspace{1mm}Adama~Nouboukpo} \\
	Department of Computer Science\\
	University of Quebec in Outaouais (UQO)\\
	Gatineau, QC J8X 3X7 \\
}
\else
\usepackage{authblk}

\newbox{\orcid}\sbox{\orcid}{\includegraphics[scale=0.06]{orcid.pdf}} 
\author[1]{%
	\href{https://orcid.org/0009-0002-7235-8653}{\usebox{\orcid}\hspace{1mm}Bappa~Muktar\thanks{\texttt{mukb06@uqo.ca}}}%
}
\author[1,2]{%
	\href{https://orcid.org/0009-0009-0944-9721}{\usebox{\orcid}\hspace{1mm}Vincent.~Fono\thanks{\texttt{stariate@ee.mount-sheikh.edu}}}%
}
\affil[1]{Department of Computer Science, Cranberry-Lemon University, Pittsburgh, PA 15213}
\affil[2]{Department of Electrical Engineering, Mount-Sheikh University, Santa Narimana, Levand}
\fi

\renewcommand{\shorttitle}{\textit{arXiv} Template}

\hypersetup{
pdftitle={A template for the arxiv style},
pdfsubject={q-bio.NC, q-bio.QM},
pdfauthor={David S.~Hippocampus, Elias D.~Striatum},
pdfkeywords={First keyword, Second keyword, More},
}

\begin{document}
\maketitle

\begin{abstract}
	Vehicular Ad Hoc Networks (VANETs) play a key role in Intelligent Transportation Systems (ITS), particularly in enabling real-time communication for emergency vehicles. However, Distributed Denial of Service (DDoS) attacks, which interfere with safety-critical communication channels, can severely impair their reliability. This study introduces a robust and scalable framework to detect DDoS attacks in highway-based VANET environments. A synthetic dataset was constructed using Network Simulator 3 (NS-3) in conjunction with the Simulation of Urban Mobility (SUMO) and further enriched with real-world mobility traces from Germany’s A81 highway, extracted via OpenStreetMap (OSM). Three traffic categories were simulated: DDoS, VoIP, and TCP-based video streaming (VideoTCP). The data preprocessing pipeline included normalization, signal-to-noise ratio (SNR) feature engineering, missing value imputation, and class balancing using the Synthetic Minority Over-sampling Technique (SMOTE). Feature importance was assessed using SHapley Additive exPlanations (SHAP). Eleven classifiers were benchmarked, among them XGBoost (XGB), CatBoost (CB), AdaBoost (AB), GradientBoosting (GB), and an Artificial Neural Network (ANN). XGB and CB achieved the best performance, each attaining an F1‑score of 96\%. These results highlight the robustness of the proposed framework and its potential for real-time deployment in VANETs to secure critical emergency communications.
\end{abstract}


\section{Introduction} \label{sect:s1}
VANETs have emerged as a cornerstone of ITS, enabling real-time communication between vehicles and infrastructure to improve traffic efficiency and road safety~\cite{ref-1, ref-2}. These systems are particularly vital for emergency response units, which rely on uninterrupted connectivity to minimize response time and save lives. However, their open communication channels, decentralized architecture, and dynamic topology expose them to a wide range of cybersecurity threats~\cite{ref-3}. Among the most critical of these threats are DDoS attacks, which aim to overwhelm network resources and degrade the performance of safety-critical services. Such disruptions can cause severe consequences, including delayed emergency interventions, increased traffic congestion, and potential loss of life~\cite{ref-4, ref-5}.

Despite increasing academic interest in intrusion detection systems for VANETs, many existing studies present notable limitations, such as exclusive reliance on synthetic datasets, lack of reproducibility, and a predominant focus on dense urban environments~\cite{ref-3, ref-6}. In particular, realistic highway scenarios—where uninterrupted communication for emergency vehicles is equally critical—remain significantly underexplored. Moreover, most prior research depends on a single machine learning classifier, which limits the robustness and generalization capacity of the proposed models.

To bridge these gaps, this paper proposes a comprehensive machine learning-based framework for detecting DDoS attacks in VANETs operating in highway environments.

The main contributions of this work are as follows:
\begin{itemize}
	\item We design and simulate realistic VANET traffic using the NS-3 and SUMO simulators, incorporating real-world vehicle mobility traces from Germany’s A81 highway extracted via OSM.
	\item We evaluate a wide range of supervised learning algorithms, including XGB, CB, AB, Extra Trees (ET), Random Forest (RF), GB, Support Vector Machine (SVM), K-Nearest Neighbors (KNN), Logistic Regression (LR), Decision Tree (DT), and ANN.
	\item We apply SHAP to assess feature importance, thereby enhancing the interpretability and reliability of the models. The proposed framework achieves excellent predictive performance, with F1-scores reaching up to 96\% for XGB and CB classifiers.
\end{itemize}

The remainder of this paper is organized as follows: Section~\ref{sect:s2}~presents a comprehensive literature review of machine learning-based intrusion detection in VANETs. Section~\ref{sect:s3}~details the methodology, including dataset generation and preprocessing. Section~\ref{sect:s4}~describes the classifiers used and the predictive modeling approach. Section~\ref{sect:s5}~reports and discusses the experimental results. Finally, Section~\ref{sect:s6}~concludes the paper and outlines future research directions.

\section{Literature Review} \label{sect:s2}
Securing VANETs against DDoS attacks has emerged as a critical research area due to the potential disruptions in vital communication channels, especially those involving emergency vehicles. Recent advances have emphasized developing robust, accurate, real-time intrusion detection mechanisms utilizing machine learning (ML) and deep learning (DL) approaches.

Several researchers have investigated innovative machine learning models tailored explicitly to the unique constraints of VANET environments. For instance, Setia et al. proposed a framework employing machine learning combined with fuzzification methods within cloud-based VANET systems, achieving a remarkable accuracy of 99.59\% in proactively detecting DDoS threats~\cite{ref-7}. Similarly, Polat, O. et al. introduced a hybrid model blending a one-dimensional Convolutional Neural Network (1D-CNN) with decision trees for real-time detection in Software-Defined Vehicular Ad-Hoc Networks (SD-VANETs), attaining an accuracy close to 90\%~\cite{ref-8}. Further expanding this direction, Polat, H. et al. presented an advanced deep learning architecture using stacked sparse autoencoders combined with a softmax classifier, significantly improving accuracy to approximately 96.9\% in SDN-based VANET scenarios~\cite{ref-9}.

Addressing not only attack detection but also network congestion, Gopi et al. developed a two-phase Intelligent DoS Attack Detection with Congestion Control (IDoS-CC) system. Their methodology combined Teaching and Learning-Based Optimization (TLBO) with a Gated Recurrent Unit (GRU) deep learning model, demonstrating substantial reductions in network congestion and improved detection accuracy~\cite{ref-10}. Kadam et al. also contributed notably by proposing a hybrid classification approach (KSVM) integrating K-Nearest Neighbors (KNN) and Support Vector Machines (SVM), exhibiting superior sensitivity, recall, and precision compared to traditional classifiers~\cite{ref-11}.

Data realism and reproducibility represent essential challenges often overlooked in the literature. In response, Alkadiri et al. generated a contemporary dataset leveraging OMNeT++, Veins, and SUMO simulations, optimized via SMOTE and classified using the XGBoost algorithm, achieving an F1-score of approximately 99\%~\cite{ref-12}. Similarly, Rashid et al. adopted OMNeT++ and SUMO for a realistic VANET simulation, presenting a real-time adaptive framework with various ML classifiers, yielding accuracies of up to 99\%~\cite{ref-13}. Anyanwu et al. further optimized detection by integrating Radial Basis Function SVM (RBF-SVM) with Grid Search Cross-Validation, showing detection rates of 99.22\% on realistic SDN-based VANET datasets~\cite{ref-14}.

Hybrid optimization and multi-stage detection systems have also been extensively explored. Marwah et al. combined modified SVM enhanced by Harris Hawks Optimization (HHO) and Whale-Dragonfly optimization for efficient routing and bandwidth allocation, significantly improving throughput and reducing communication overhead under DDoS conditions~\cite{ref-15}. Adhikary et al. developed a hybrid model merging AnovaDot and RBFDot SVM kernels into a chained detection mechanism, achieving improved robustness and detection accuracy compared to single-kernel models~\cite{ref-16}. Moreover, Tariq et al. proposed a comprehensive detection framework integrating Autoencoders, LSTM, clustering methods, fog computing, and blockchain technology, offering a low-latency, scalable, and robust solution with a detection rate of approximately 94\%~\cite{ref-17}.

Deep learning-based anomaly detection approaches have recently gained momentum due to their scalability and superior pattern recognition capabilities. Lekshmi et al. leveraged convolutional autoencoders coupled with LSTM networks and self-attention mechanisms, achieving an F1-score of 98.20\% in detecting DDoS attacks on realistic VANET data~\cite{ref-18}. Similarly, Haydari et al. introduced a semi-supervised, non-parametric intrusion detection system using roadside units (RSUs), capable of detecting novel attack patterns without prior knowledge, significantly enhancing real-time responsiveness and detection accuracy~\cite{ref-19}.

While extensive progress has been made, gaps remain in terms of evaluating these methodologies in realistic highway scenarios. Most existing works predominantly target dense urban environments or lack reproducible real-world mobility data, limiting the generalizability of results. Additionally, comprehensive comparisons of various machine learning classifiers within a unified, realistic highway scenario remain scarce.

Our study aims to address these critical gaps by evaluating multiple prominent ML classifiers—including XGB, CB, AB, ET, RF, GB, SVM, KNN, LR, DT, and ANN—in a realistic VANET highway scenario. Leveraging NS-3 and SUMO simulators enriched with real mobility data from the A81 highway in Germany, our approach not only ensures realism but also enables reproducibility. Furthermore, data balancing through SMOTE and rigorous performance evaluation metrics (accuracy, precision, recall, and F1-score) strengthen our methodological framework, providing a robust and comprehensive assessment of classifier effectiveness.

Table~\ref{tabref:table-1}~below summarizes and positions our work compared to existing state-of-the-art approaches based on several critical criteria.

\begin{longtable}{C{1.5cm} C{2.5cm} C{3.2cm} C{1.5cm} C{2.5cm} C{1.8cm} C{3.0cm}}
\caption{Comparative summary of DDoS detection in VANETs (continued on next page)} \label{tabref:table-1} \\
\toprule
\textbf{Reference} & \textbf{Model Type} & \textbf{Simulation Environment} & \textbf{Attack Type} & \textbf{Detection Approach} & \textbf{Data Balancing} & \textbf{Best Reported Metric} \\
\midrule
\endfirsthead

\multicolumn{7}{c}%
{{\bfseries Table~\thetable\ Continued from previous page}} \\
\toprule
\textbf{Reference} & \textbf{Model Type} & \textbf{Simulation Environment} & \textbf{Attack Type} & \textbf{Detection Approach} & \textbf{Data Balancing} & \textbf{Best Reported Metric} \\
\midrule
\endhead

\bottomrule
\multicolumn{7}{r}{{Continued on next page}} \\
\endfoot

\bottomrule
\endlastfoot

\cite{ref-7}  & ML + Fuzzification           & NS-2 (VANET cloud sim.)            & DDoS   & Fuzzy logic aided ML classifier     & None  & Accuracy: 99.59\% \\
\cite{ref-8}  & 1D-CNN + Decision Tree       & SD-VANET (Mininet+SUMO)            & DDoS   & Hybrid CNN+DT classification        & None  & $\sim$90\% accuracy \\
\cite{ref-9}  & Stacked Autoencoder (SSAE)   & SDN-based VANET (sim.)             & DDoS   & Deep learning (SSAE + Softmax)      & None  & Accuracy: 96.9\% \\
\cite{ref-10} & TLBO + GRU (two-stage)       & VANET traffic simulation           & DoS    & Optimization (TLBO) + RNN classifier& None  & Not specified \\
\cite{ref-11} & KNN + SVM (Hybrid KSVM)      & Simulated VANET (not spec.)        & DDoS   & Combined KNN/SVM classifier         & None  & Accuracy: 92.46\% \\
\cite{ref-12} & XGBoost                      & OMNeT++/Veins + SUMO               & DDoS   & Supervised ML (tree-based)          & SMOTE & F1-score $\approx$99\% \\
\cite{ref-13} & Multi-ML ensemble            & OMNeT++/Veins + SUMO               & DDoS   & Distributed multi-layer IDS         & None  & Accuracy up to 99\% \\
\cite{ref-14} & RBF-SVM (optimized)          & SDN-VANET (realistic data)         & DDoS   & SVM + grid-search tuning            & None  & Detection Rate: 99.22\% \\
\cite{ref-15} & SVM + HHO + WDO              & Simulated VANET (Hwy)              & DDoS   & Optimized SVM (HHO, Whale-Dragonfly)& None  & F1-score: 96\% \\
\cite{ref-16} & Dual-kernel SVM              & Simulated VANET (RSUs)             & DDoS   & Chained AnovaDot+RBF SVM            & None  & Accuracy $\sim$96–98\% \\
\cite{ref-17} & Autoenc. + LSTM + BC (fog)   & Simulated SD-VANET (fog)           & DDoS   & Hybrid IDS + blockchain             & None  & Detection Rate $\approx$94\% \\
\cite{ref-18} & ConvAE + LSTM                & Simulated VANET data               & DDoS   & Deep anomaly detection              & None  & F1-score: 98.20\% \\
\cite{ref-19} & Statistical (non-param.)     & SUMO + real traffic traces         & DDoS   & RSU-based anomaly detection         & None  & Detection $\sim$94\% \\
\textbf{Ours} & ML \& DL (XGB, CB, ANN...)   & NS-3, SUMO, Real traces (A81 Hwy)  & DDoS   & ML/DL classifiers (with SMOTE)      & SMOTE & F1-score $\sim$96\% \\
\end{longtable}

This comparative analysis underscores the novelty and relevance of our research, emphasizing both methodological rigor and practical applicability, thus effectively filling the identified gaps in the current state of VANET cybersecurity research.

\section{Methodology} \label{sect:s3}
This section outlines the methodological framework for developing a robust classification model for DDoS attacks in a VANET environment, simulating a realistic highway scenario.

\subsection{Experimental Architecture}
This section outlines the architecture and methodology used to simulate a realistic highway-based VANET under coordinated DDoS attacks. It details the scenario design, simulator integration, and incorporation of real mobility traces to ensure data realism and model applicability.

\subsubsection{Scenario Description}
The data collection scenario is structured to simulate a VANET highway environment with 13 vehicles (from V$_0$ to V$_{12}$) moving at a constant speed. V$_0$ to V$_2$ act as legitimate nodes, while V$_3$ to V$_{12}$ act as malicious nodes. Vehicle V$_0$, which symbolizes an emergency vehicle (for instance, a police car), will generate TCP traffic to vehicle V$_2$, which simulates a real-time video streaming application. At the same time, vehicle V$_1$ is transmitting VoIP messages over UDP to the same destination. On the other hand, malicious nodes (V$_3$ to V$_{12}$) initiated a DDoS attack by overwhelming V$_2$ with high UDP traffic flows to disrupt its communication capabilities. This scenario demonstrates the critical security threat in a VANET highway environment, where a coordinated cyberattack threatens the emergency vehicle's operational integrity. Table~\ref{tabref:table-2} presents the NS-3 simulation parameters used in the VANET DDoS scenario.

\begin{table}[h!]
\centering
\caption{Simulation parameters used for the VANET DDoS scenario}
\label{tabref:table-2}
\begin{tabular}{ll}
\hline
\textbf{Parameter} & \textbf{Value / Description} \\
\hline
Simulation Time & 30 seconds \\
Number of Nodes & 13 vehicles total (3 legitimate, 10 malicious) \\
Legitimate Vehicles & V\textsubscript{0}: TCP (Video), V\textsubscript{1}: UDP (VoIP), V\textsubscript{2}: Sink \\
Malicious Vehicles & V\textsubscript{3} to V\textsubscript{12} (UDP DDoS) \\
WiFi Standard & IEEE 802.11 (10 MHz channel bandwidth) \\
WiFi Range & 250 meters \\
Routing Protocol & OLSR (Optimized Link State Routing) \\
Propagation Model & Two-Ray Ground Propagation Loss Model \\
Mobility Model & Ns2MobilityHelper (A81 highway traces) \\
Packet Size (VoIP) & 160 bytes \\
VoIP Rate & 64 Kbps \\
Packet Size (DDoS) & 1024 bytes \\
DDoS Rate per Bot & 1 Mbps \\
Traffic Classification & TCP (Video), UDP (VoIP/DDoS), based on source address \\
Monitoring Tools & FlowMonitor (with SNR via MonitorSniffRx) \\
Output Files & \texttt{vanet-ddos-data.csv}, \texttt{vanet-ddos-flowmon.xml} \\
\hline
\end{tabular}
\end{table}

\subsubsection{NS-3 and SUMO Integration}
The experiment uses NS-3~\cite{ref-20}~and SUMO~\cite{ref-21}~simulators to simulate communication protocols and vehicle dynamics. NS-3 handles network stack, protocol behavior, and traffic generation, while SUMO provides the precise mobility dynamics of the vehicle for realistic traffic scenarios.

\subsubsection{Incorporation of Real Mobility Traces}
\begin{figure}[H]
    \centering
    \begin{subfigure}[t]{0.3\textwidth}
        \centering
        \includegraphics[width=\linewidth]{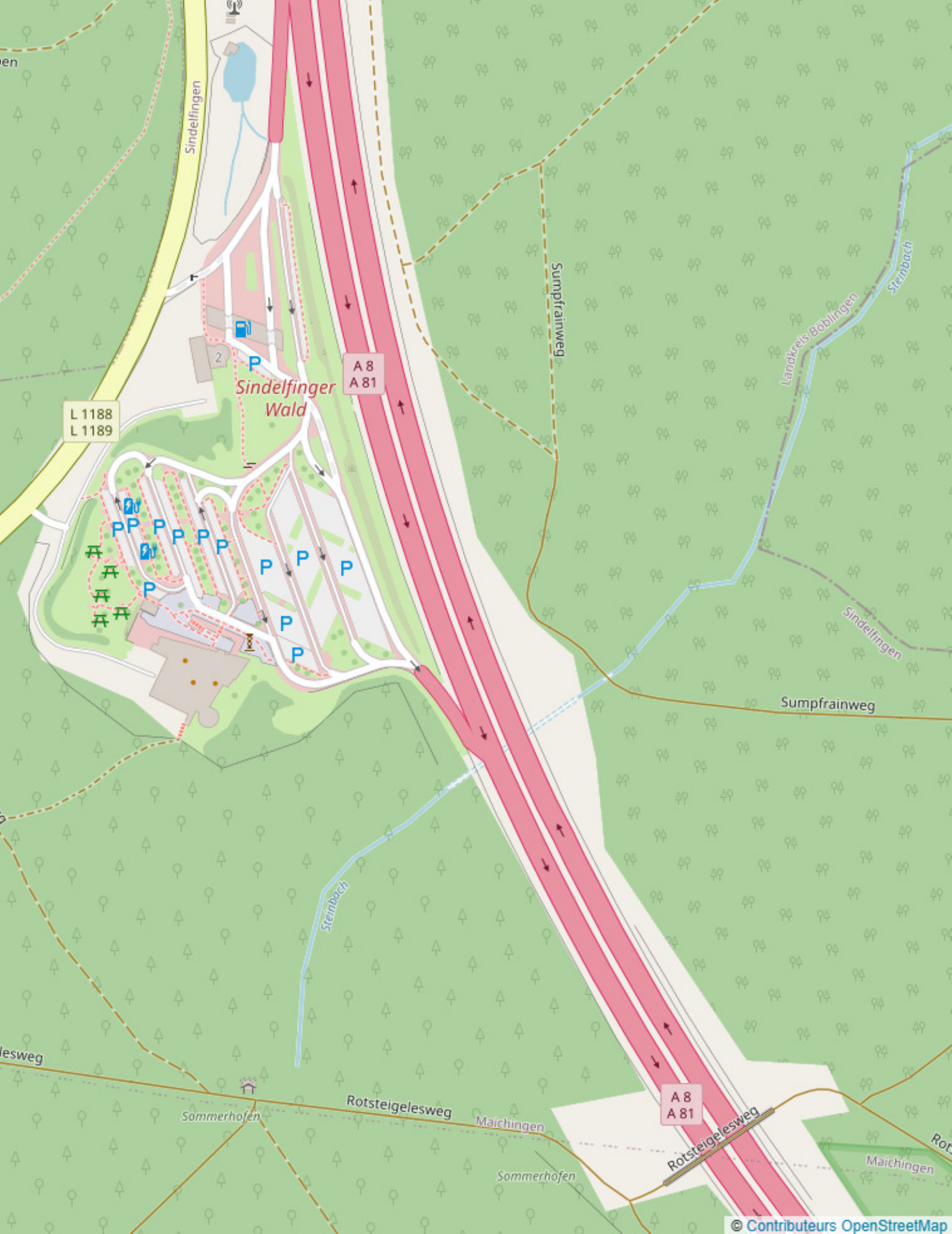}
        \caption{A81 highway segment extracted from OpenStreetMap.}
        \label{fig:a81-osm}
    \end{subfigure}
    \hfill
    \begin{subfigure}[t]{0.65\textwidth}
        \centering
        \includegraphics[width=\linewidth]{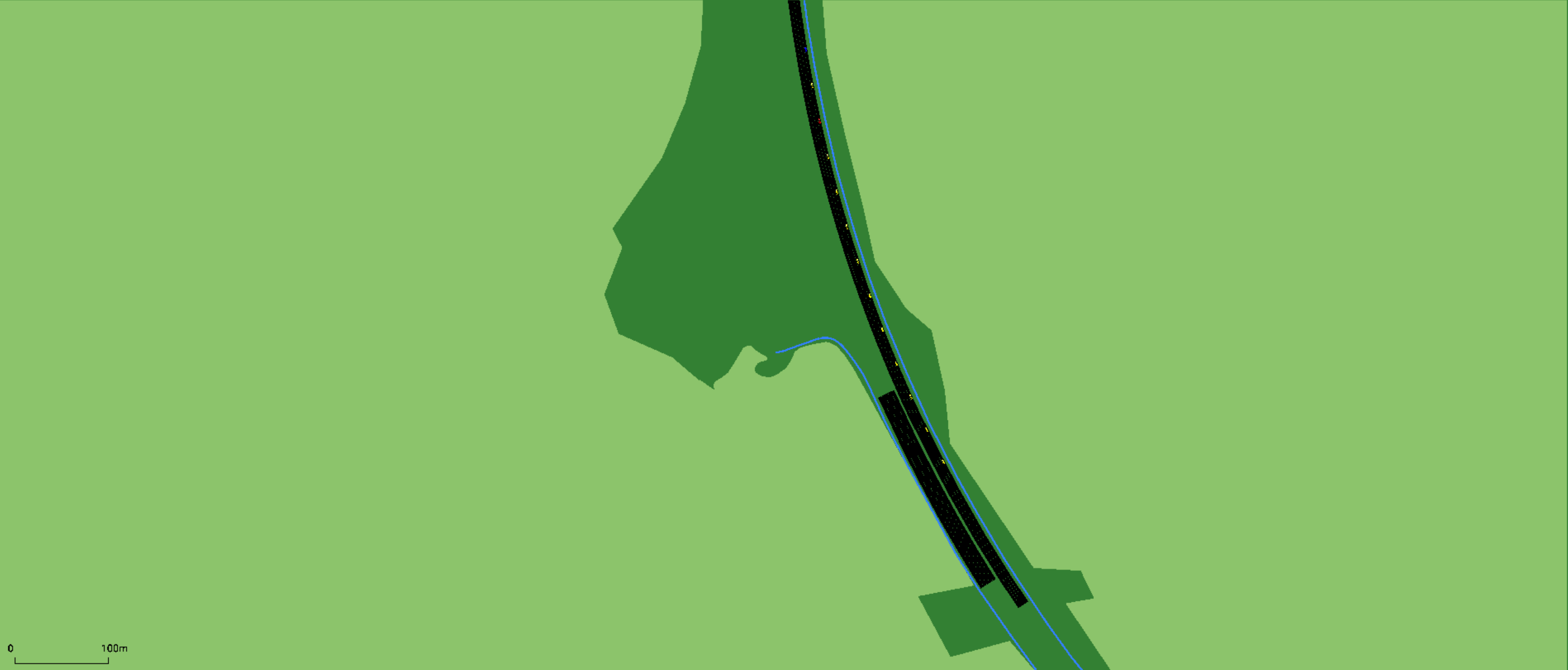}
        \caption{Imported road segment visualized in SUMO.}
        \label{fig:a81-sumo}
    \end{subfigure}
    \caption{Visualization of the A81 highway segment used in the simulation. (a) Map segment from OSM. (b) Simulation rendering in SUMO.}
    \label{fig:fig-1}
\end{figure}

To further enhance the realism of the simulation, real-world mobility traces from the A81 highway in Germany were integrated into the SUMO simulation and imported into NS-3 using the Ns2MobilityHelper module. This integration ensures that the generated dataset reflects authentic vehicular behavior and spatial-temporal patterns, thus increasing the applicability and reliability of the intrusion detection model trained on this data. Figure~\ref{fig:fig-1} illustrates the A81 highway in OSM and its corresponding import within the SUMO environment.

\subsection{Data Generation and Labeling}

The simulated dataset utilized in this study comprises three distinct classes of network traffic: \textit{(DDoS)}, \textit{Voice over IP (VoIP)}, and \textit{VideoTCP}. Each traffic category was generated using appropriate application models within the NS-3 simulation environment. Specifically, \textit{VideoTCP} traffic, emulating a real-time video streaming application, was produced using the \texttt{BulkSendHelper} application over a TCP connection directed toward the target vehicle. Concurrently, \textit{VoIP} traffic was simulated using the \texttt{OnOffHelper} application, configured at a constant data rate of $64\,\text{kbps}$ and a fixed packet size of $160$ bytes, thereby adhering to the widely used G.711 standard in VoIP communications. In contrast, \textit{DDoS} traffic was generated using the same \texttt{OnOffHelper} application, but set to a significantly higher data rate of $1\,\text{Mbps}$ per flow, explicitly modeling malicious traffic intended to saturate network resources.

To characterize the behavior and performance of each network flow, several relevant metrics were collected using the \texttt{FlowMonitor} module in NS-3. Key metrics extracted include the average throughput, measured in kilobits per second (kbps), computed according to the following equation:
\[
\text{Throughput} = \frac{8 \times \text{RxBytes}}{\text{FlowDuration}\times 10^3}
\]
Where \textit{RxBytes} denotes the total number of bytes received and \textit{FlowDuration} represents the effective duration of the flow in seconds. The mean delay was calculated using:
\[
\text{MeanDelay} = \frac{\sum_{i=1}^{N_{rx}}\text{Delay}_i}{N_{rx}}
\]
Where $\text{Delay}_i$ is the delay experienced by each successfully received packet and $N_{rx}$ corresponds to the total number of received packets. Additionally, the packet loss rate (\textit{LostPackets}) was determined by calculating the difference between transmitted ($N_{tx}$) and received ($N_{rx}$) packets:
\[
\text{LostPackets} = N_{tx} - N_{rx}
\]

\begin{figure}[H]
    \centering
    \includegraphics[width=0.95\textwidth]{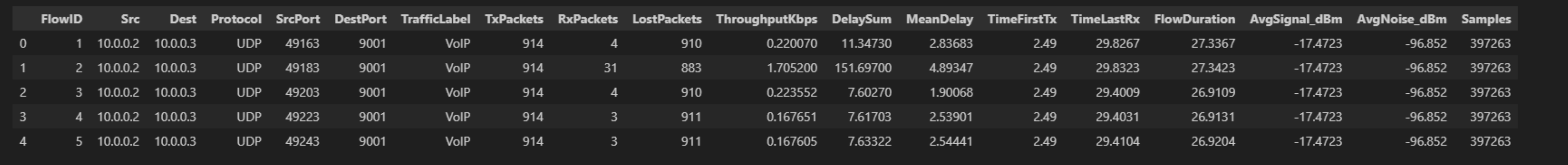}
    \caption{Dataset sample}
    \label{fig:fig2}
\end{figure}

Lastly, each network flow was explicitly labeled according to its traffic class (\textit{DDoS}, \textit{VoIP}, or \textit{VideoTCP}) based on the originating IP address and the employed network protocol. Consequently, TCP-based flows were systematically classified as \textit{VideoTCP}, UDP-based flows originating from legitimate nodes (IP addresses $\leq$ 10.0.0.3) were labeled as \textit{VoIP}, whereas UDP flows initiated by malicious bot nodes were categorized as \textit{DDoS}. This meticulous labeling procedure enhances the reliability and accuracy of the dataset, facilitating the development of robust and effective intrusion detection models. Figure~\ref{fig:fig2} shows the first five rows of the dataset sample extracted from the NS-3 simulation.

\subsection{Data Preprocessing}

The preprocessing stage is a fundamental step in building an effective intrusion detection model. This process was structured into three main phases: data cleaning and normalization, creation of a derived SNR variable, and class rebalancing through oversampling techniques.

\subsubsection{Cleaning and Normalization}

The raw dataset initially consisted of 6882 network flows described by 19 features, including identifiers, traffic characteristics, performance metrics, and physical measurements such as average signal and noise power. Several cleaning operations were applied:

\begin{itemize}
    \item Removal of non-informative or highly correlated features: Columns such as \texttt{FlowID}, \texttt{Src}, \texttt{Dest}, \texttt{SrcPort}, \texttt{DestPort}, and \texttt{Samples} were discarded due to their low predictive value. Similarly, the temporal features \texttt{TimeFirstTx} and \texttt{TimeLastRx} were removed in favor of the derived feature \texttt{FlowDuration}, and \texttt{DelaySum} was excluded in favor of \texttt{MeanDelay}.
    \item Categorical feature encoding: The categorical variables \texttt{Protocol} and \texttt{TrafficLabel} were converted to numerical representations using \texttt{LabelEncoder}, where \texttt{DDoS}, \texttt{VoIP}, and \texttt{VideoTCP} were encoded as 0, 2, and 1, respectively.
    \item Duplicate removal: Approximately 7.5\% of the data were identified as duplicates and subsequently removed to reduce model bias.
    \item Normalization: All numerical features were normalized using \texttt{StandardScaler} to enforce zero mean and unit variance—an essential condition for many machine learning algorithms.
\end{itemize}

\subsubsection{SNR Feature Engineering}

Although the dataset initially contained the fields \texttt{AvgSignal\_dBm} and \texttt{AvgNoise\_dBm}, a new variable representing the average Signal-to-Noise Ratio (SNR) was computed as follows:

\[
\overline{\text{SNR}} = \overline{S} - \overline{N}
\]

where \( \overline{S} \) and \( \overline{N} \) denote the mean received signal and noise power respectively, measured in dBm. However, SHAP (SHapley Additive exPlanations) analysis revealed that these features had negligible predictive value in the highway VANET scenario, and they were therefore excluded from the final dataset used for training.

\subsubsection{Class Rebalancing Using SMOTE}
\label{sect:s333}

\begin{figure}[H]
    \centering
    \includegraphics[width=0.5\textwidth]{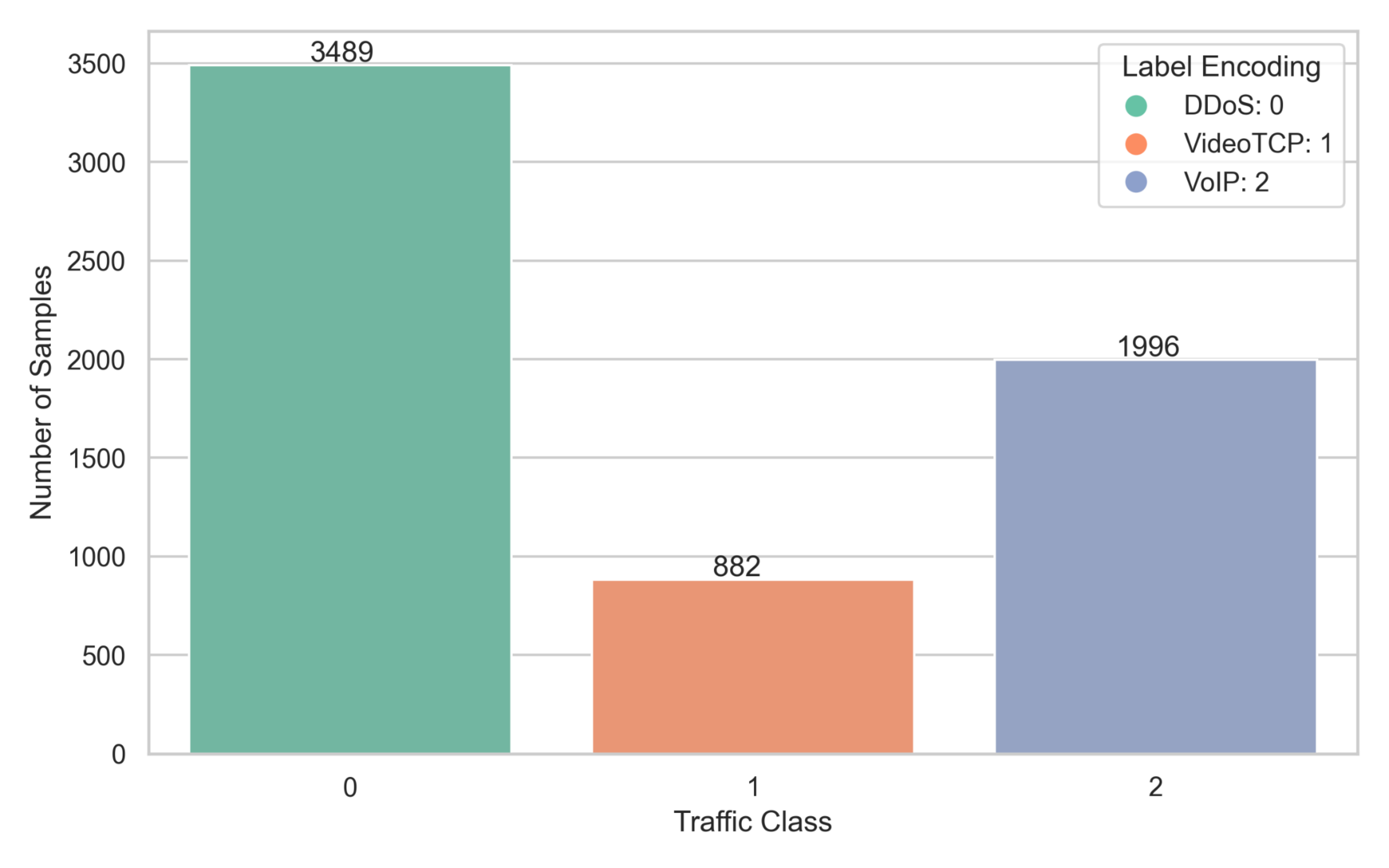}
    \caption{Traffic label distribution before SMOTE}
    \label{fig:fig3}
\end{figure}

The Figure~\ref{fig:fig3} below highlights a significant class imbalance: 3489 \texttt{DDoS} flows, 1996 \texttt{VoIP} flows, and only 882 \texttt{VideoTCP} flows. To address this, we applied the SMOTE~\cite{ref-22}~to the training data. SMOTE generates synthetic samples for the minority classes, resulting in a balanced training set with 2617 flows per class.

This rebalancing significantly improved model generalization and reduced bias toward the majority class during training.

\subsection{Feature Selection}

\begin{figure}[H]
    \centering
    \includegraphics[width=0.8\textwidth]{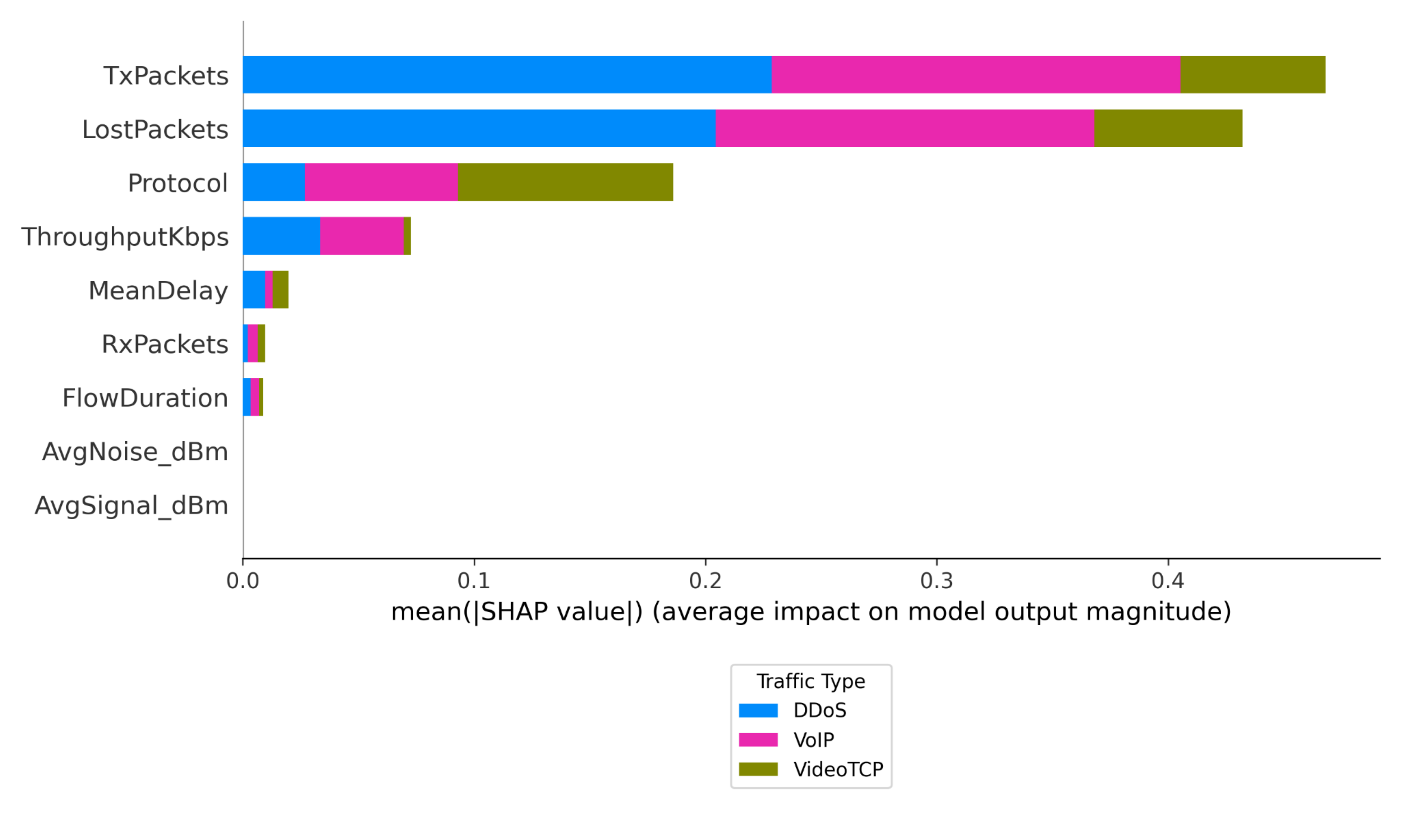}
    \caption{Feature importance based on SHAP values}
    \label{fig:fig-4}
\end{figure}

Feature selection plays a pivotal role in the development of any predictive model, particularly in the context of VANETs, where the dataset may include redundant or highly correlated variables. To identify the most relevant attributes for classifying network traffic (\textit{DDoS}, \textit{VoIP}, and \textit{VideoTCP}), we adopted an interpretability-based approach using SHAP values (see Fig.~\ref{fig:fig-4}). This method quantifies the marginal contribution of each feature to the model's output while accounting for complex interdependencies among features.

As illustrated in Fig.~\ref{fig:fig-4}, the SHAP analysis highlighted \texttt{TxPackets}, \texttt{LostPackets}, and \texttt{Protocol} as the most influential features in predicting the traffic class. Although these features exhibit some degree of correlation, they offer complementary insights into traffic intensity and anomalous behavior, such as packet losses resulting from DDoS attacks.

Nonetheless, \texttt{TxPackets} and \texttt{LostPackets}, despite their high SHAP scores and strong correlation with the target variable, were deliberately excluded from the final feature set to mitigate multicollinearity effects. These variables directly influence several other performance metrics (e.g., \texttt{ThroughputKbps} and \texttt{MeanDelay}), and including them could introduce bias by over-representing certain aspects of the traffic.

The final selection includes the following features:
\begin{itemize}
    \item Protocol: distinguishes UDP flows (VoIP) from TCP flows (VideoTCP), and supports the identification of traffic patterns typical of DDoS attacks.
    \item ThroughputKbps: reflects traffic intensity and helps discriminate between high-volume flows such as those generated by \textit{VideoTCP} and \textit{DDoS}.
    \item MeanDelay: captures average packet latency, which is critical for detecting delays caused by attacks or real-time services like VoIP.
    \item RxPackets: although moderately ranked in SHAP importance, this feature complements flow-level analysis without the redundancy of \texttt{TxPackets}.
    \item FlowDuration: captures the temporal dynamics of each flow and effectively substitutes highly correlated variables such as \texttt{TimeFirstTx} and \texttt{TimeLastRx}.
\end{itemize}

This refined feature set was selected based on its discriminative power while minimizing redundancy. It ensures improved robustness and interpretability of the classification model, which is essential for reliable intrusion detection in VANET environments.

\section{Modeling and Classification}\label{sect:s4}
This section presents the modeling approach to classify network traffic in a VANET scenario under DDoS conditions.

\subsection{Tested Machine Learning Models}

To assess the ability to classify network traffic in a VANET environment, several machine learning algorithms were tested, encompassing both traditional methods and more advanced ensemble and boosting techniques.

The traditional models evaluated include:
\begin{itemize}
    \item Random Forest: An ensemble method based on building multiple decision trees and averaging their predictions to improve generalization.
    \item Extra Trees: Similar to Random Forest, but introducing greater randomness in the selection of splitting thresholds to enhance diversity.
    \item Decision Tree: A simple hierarchical model based on attribute-based decision rules.
    \item Logistic Regression: A linear model adapted for multiclass classification through the softmax activation function.
    \item Support Vector Machine: Using an optimized linear kernel to separate network traffic classes effectively.
    \item K-Nearest Neighbors: A non-parametric method that classifies each observation based on the majority vote among its $k$ nearest neighbors.
\end{itemize}

Advanced boosting and ensemble methods were also evaluated:
\begin{itemize}
    \item XGBoost: A gradient boosting framework optimized for multiclass classification tasks using the \texttt{multi:softmax} objective function.
    \item CatBoost: Designed to efficiently handle categorical variables and exhibit robustness against class imbalance.
    \item AdaBoost: An iterative ensemble technique that sequentially improves weak classifiers.
    \item Gradient Boosting: Builds models sequentially to correct errors made by prior models.
\end{itemize}

Finally, an Artificial Neural Network was designed and implemented using Keras. The architecture consists of:
\begin{itemize}
    \item An input layer receiving 5 features (Protocol, ThroughputKbps, MeanDelay, RxPackets, FlowDuration).
    \item A first dense hidden layer with 32 neurons and a \texttt{ReLU} activation function.
    \item A second dense hidden layer with 16 neurons, also activated by \texttt{ReLU}.
    \item A Dropout layer with a rate of $30\%$ applied after the second hidden layer to mitigate overfitting.
    \item A third dense hidden layer with 8 neurons and a \texttt{ReLU} activation function.
    \item An output dense layer with 3 neurons using the \texttt{Softmax} activation function to classify among three classes: DDoS, VoIP, and VideoTCP.
\end{itemize}

\begin{figure}[H]
    \centering
    \includegraphics[width=1.0\textwidth]{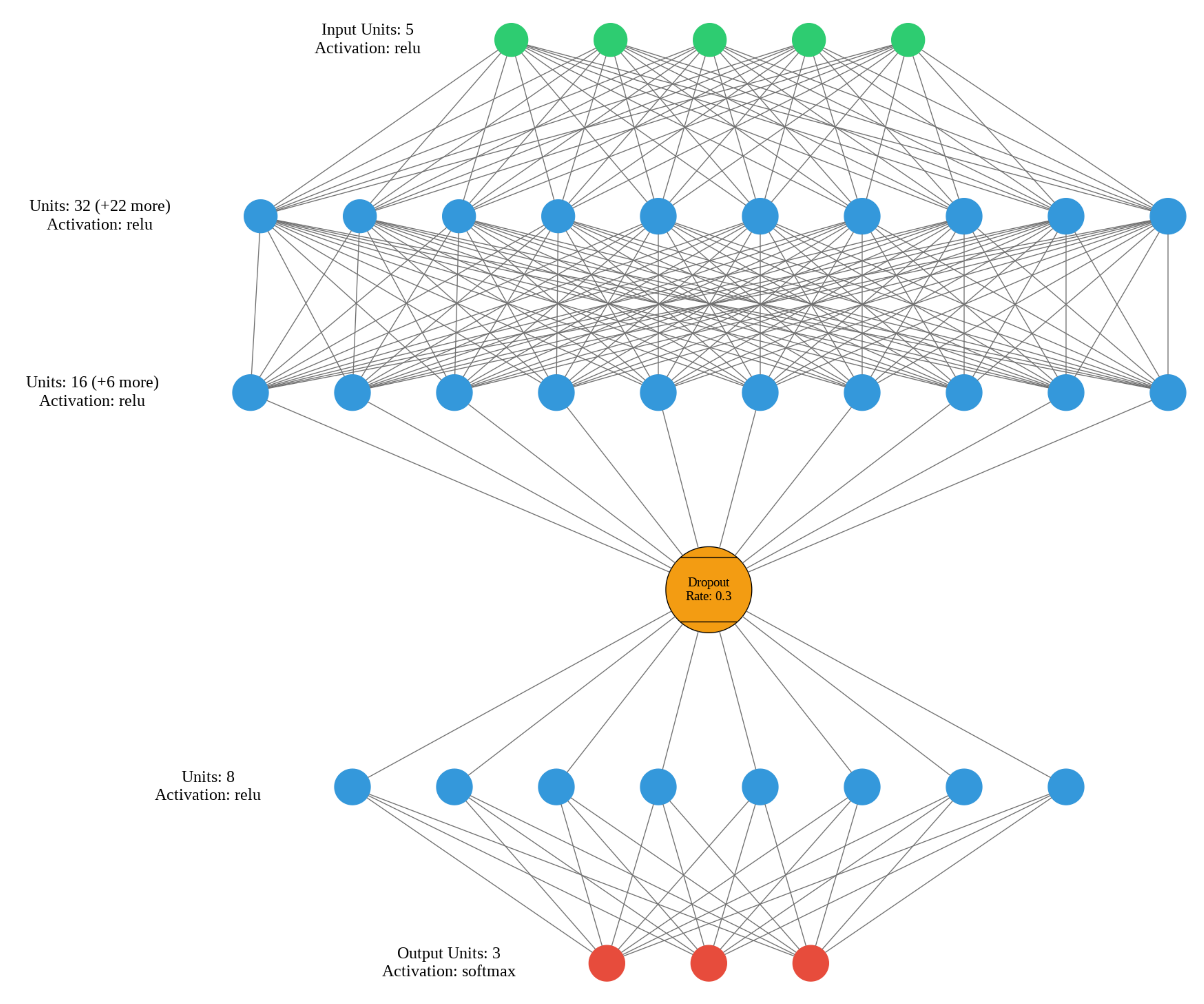}
    \caption{Architecture of the designed ANN}
    \label{fig:fig-5}
\end{figure}

Figure~\ref{fig:fig-5} illustrates the architecture of the designed ANN.

\subsection{Training and Validation}

The dataset was split into a training set ($75\%$) and a test set ($25\%$) while maintaining class proportions through a \textit{stratified split}. To address the class imbalance—particularly the under-representation of \textit{VideoTCP} traffic—the SMOTE (refer to subsection~\ref{sect:s333}) was applied to the training set, ensuring a balanced number of samples across classes.

For the scikit-learn models, training was performed after standardizing the variables using a \texttt{StandardScaler}. No explicit \texttt{class\_weight} parameter was specified since SMOTE effectively mitigated the initial class imbalance.

\begin{figure}[H] 
\centering
\begin{subfigure}[t]{0.5\textwidth}
    \centering
    \includegraphics[width=\linewidth]{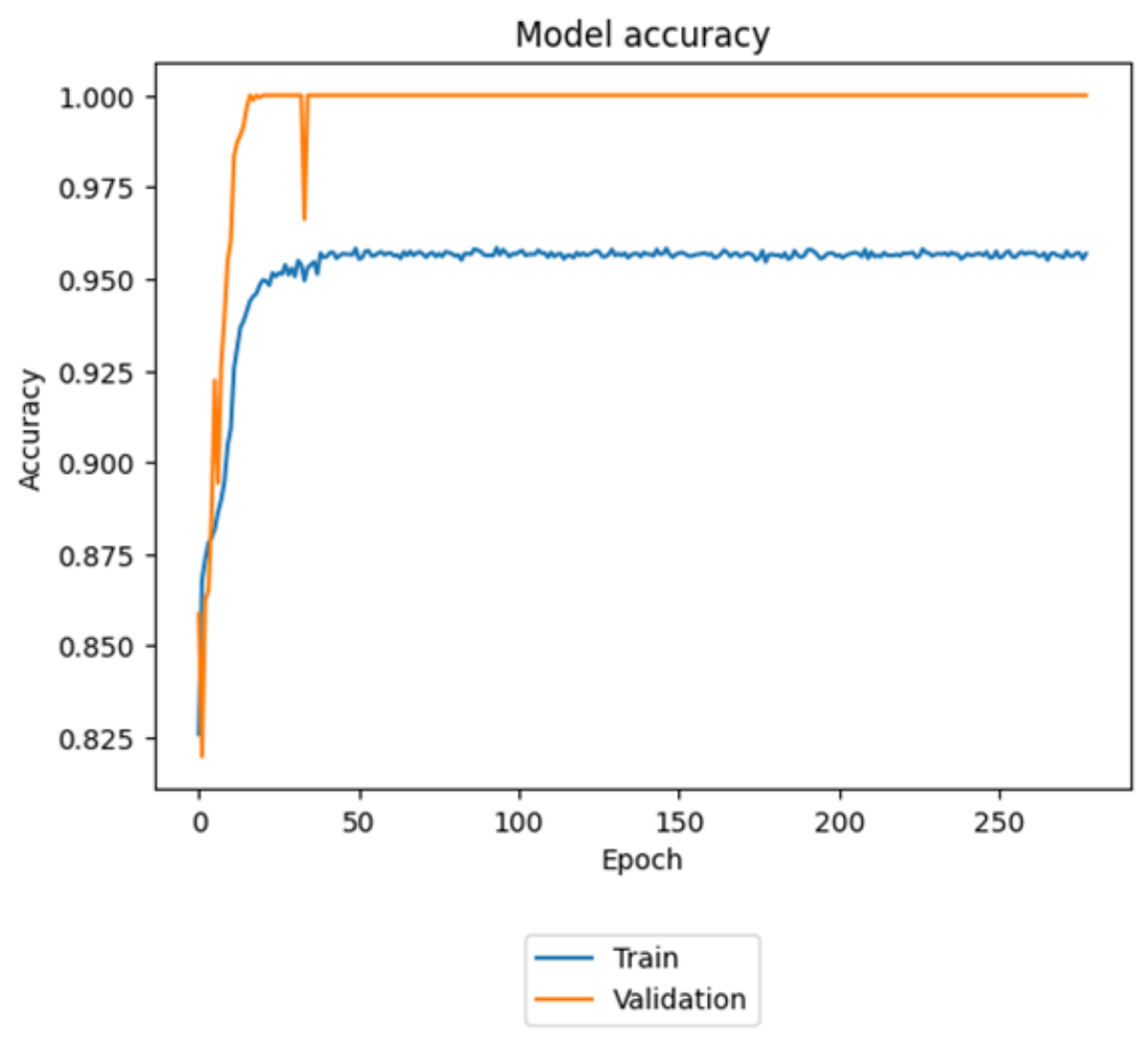}
    \caption{Evolution of model accuracy during training.}
    \label{fig:model-accuracy}
\end{subfigure}

\vspace{0.5cm}

\begin{subfigure}[t]{0.5\textwidth}
    \centering
    \includegraphics[width=\linewidth]{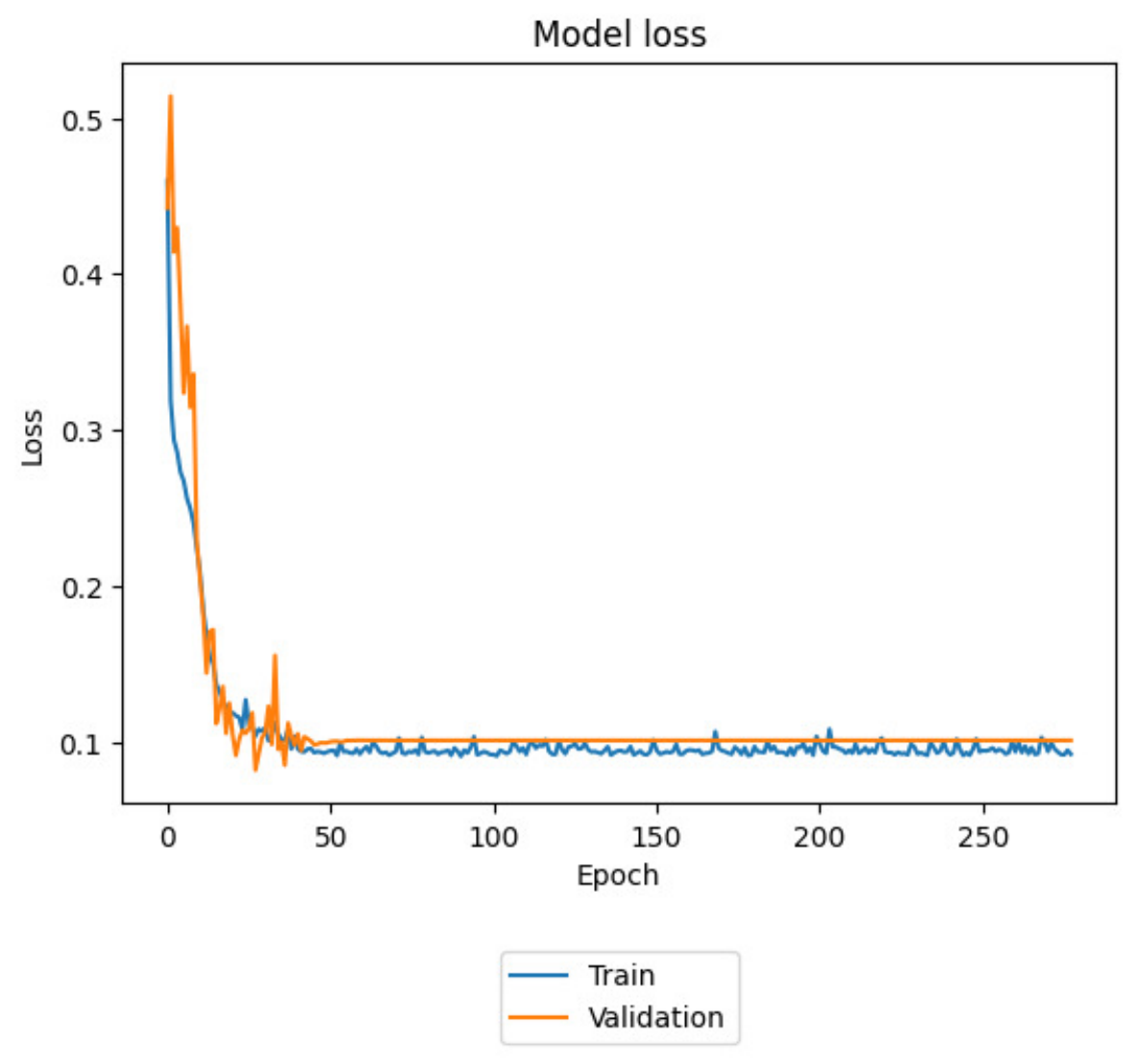}
    \caption{Evolution of model loss during training.}
    \label{fig:model-loss}
\end{subfigure}

\caption{Training history of the ANN: (a) model accuracy and (b) model loss.}
\label{fig:fig-6}
\end{figure}

For the ANN, class labels were converted into one-hot encoding before training. Validation was conducted through an internal split ($20\%$ of the training set) combined with an EarlyStopping strategy, monitoring the minimization of the validation loss. Figure~\ref{fig:fig-6} illustrates the evolution of the model's performance during training, showing (a) the model accuracy and (b) the model loss.

\subsection{Model Evaluation} 

The performance of each classification algorithm was assessed using standard evaluation metrics derived from the confusion matrix (CM), namely Accuracy, Precision, Recall, and F1-score. These metrics quantify the models’ ability to correctly classify the network traffic into the three categories: \textit{DDoS}, \textit{VoIP}, and \textit{VideoTCP}. The definitions and formulas are as follows:

\begin{itemize}
    \item \textbf{Accuracy (AC)}: Represents the ratio of correctly predicted instances over the total number of samples. It is computed using:
    \[
    Accuracy(AC) = \frac{TP + TN}{TP + TN + FP + FN}
    \]
    
    \item \textbf{Recall (R)}: Measures the proportion of true positives detected among all actual positive cases. The formula is:
    \[
    Recall(R) = \frac{TP}{TP + FN}
    \]
    
    \item \textbf{Precision (P)}: Indicates the ratio of correctly predicted positive observations to the total predicted positives:
    \[
    Precision(P) = \frac{TP}{TP + FP}
    \]

    \item \textbf{F1-score}: Combines precision and recall into a single metric by calculating their harmonic mean:
    \[
    F1\ Score = 2 \times \frac{Precision \times Recall}{Precision + Recall}
    \]
\end{itemize}

To compute these metrics for each algorithm, the confusion matrices were extracted after testing on the evaluation set. These matrices contain the number of true positive (TP), false positive (FP), true negative (TN), and false negative (FN) predictions for each class. The values were used to assess how each model performed in distinguishing between normal traffic (VoIP, VideoTCP) and malicious traffic (DDoS).

\begin{figure}[H]
    \centering
    \includegraphics[width=0.6\textwidth]{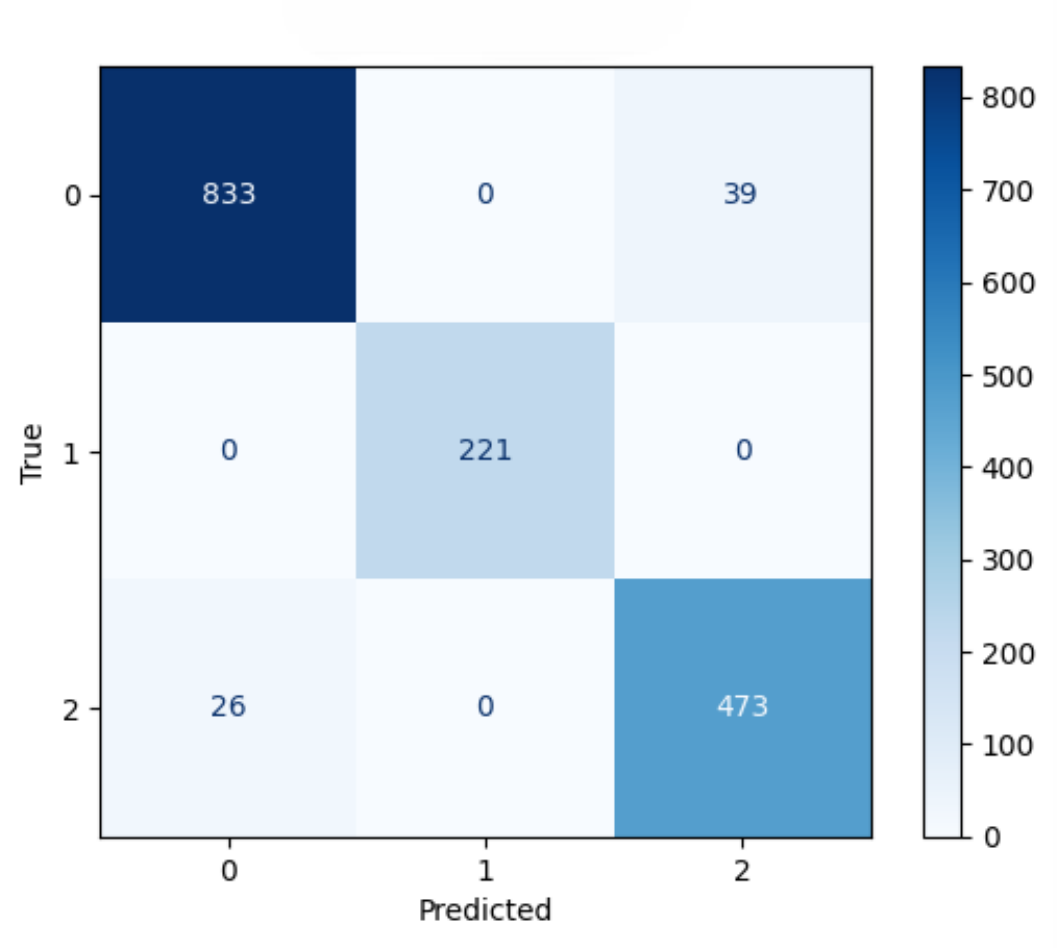}
    \caption{Confusion matrix of the XGBoost model. Class labels 0, 1, and 2 correspond to DDoS, VideoTCP, and VoIP, respectively.}
    \label{fig:fig-7}
\end{figure}

Figure~\ref{fig:fig-7} illustrates an example confusion matrix for the best-performing model (XGBoost), showing a high rate of correct predictions across all classes. This model achieved an F1-score of 0.96, with a balanced performance across the three traffic types.

\section{Results and Discussion} \label{sect:s5}

This section outlines the performance outcomes of the machine learning models used in this study and provides a corresponding analysis and interpretation of these findings.

\subsection{Results}
\begin{figure}[H]
    \centering
    \includegraphics[width=0.9\textwidth]{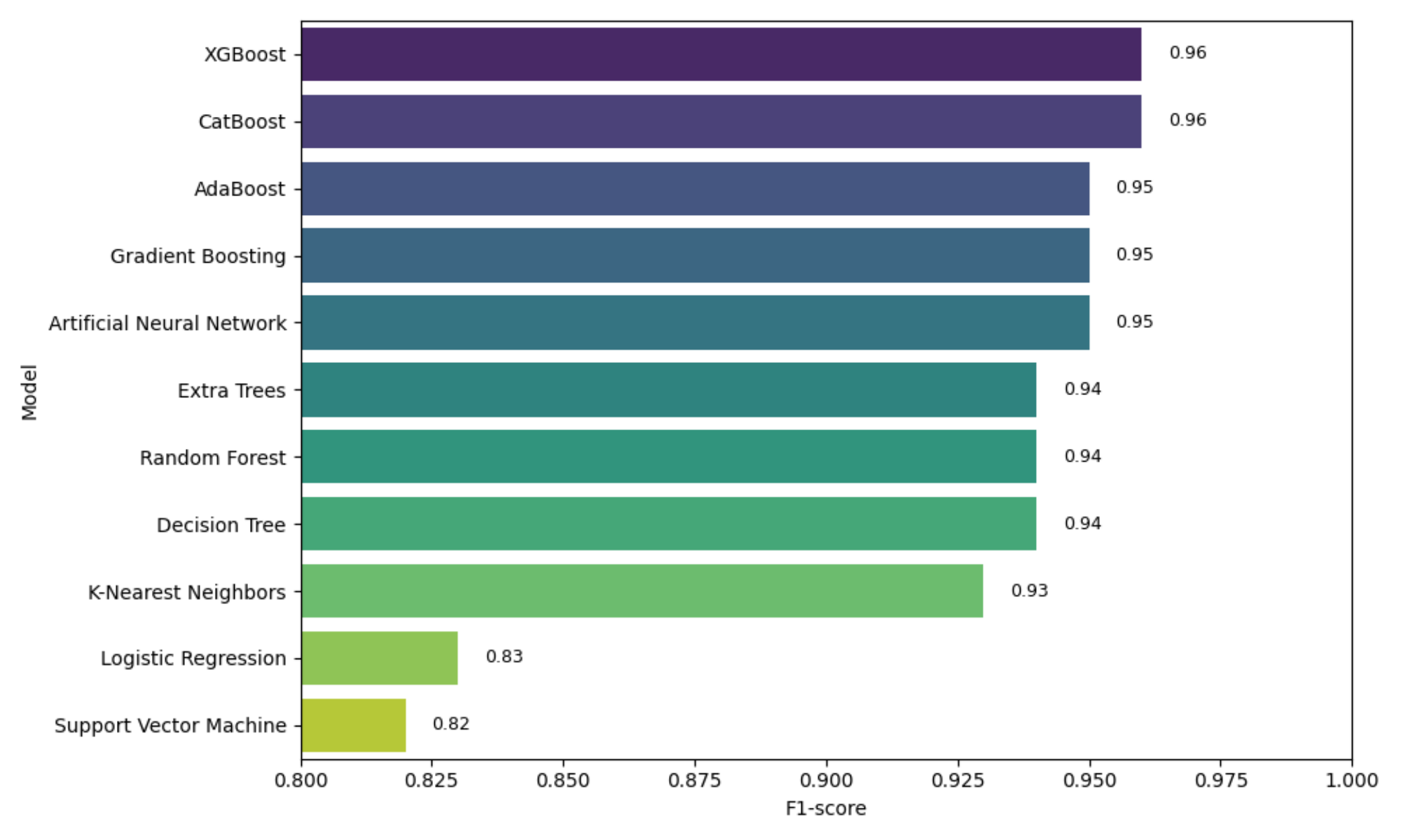}
    \caption{F1-score comparison across models.}
    \label{fig:fig-8}
\end{figure}

The classification report summary (Table~\ref{tabref:table-3}), together with the comparative analysis of F1-scores across various algorithms (Figure~\ref{fig:fig-8}), offers a thorough evaluation of the predictive capabilities of each model.

\begin{longtable}{lccccc}
\caption{Summary of classification results across different models.}\label{tabref:table-3}\\
\toprule
\textbf{Class} & \textbf{Precision} & \textbf{Recall} & \textbf{F1-score} & \textbf{Accuracy} & \textbf{Support} \\
\midrule
\endfirsthead
\multicolumn{6}{c}{\tablename\ \thetable{} -- continued from previous page} \\
\toprule
\textbf{Class} & \textbf{Precision} & \textbf{Recall} & \textbf{F1-score} & \textbf{Accuracy} & \textbf{Support} \\
\midrule
\endhead
\midrule \multicolumn{6}{r}{{Continued on next page}} \\
\endfoot
\bottomrule
\endlastfoot

\multicolumn{6}{c}{\textbf{XGBoost}} \\
\midrule
DDoS     & 0.97 & 0.96 & 0.96 & - &  872 \\
VoIP     & 1.00 & 1.00 & 1.00 & - & 221 \\
VideoTCP & 0.92 & 0.95 & 0.94 & - & 499 \\
\textbf{Overall} & \textbf{0.96} & \textbf{0.96} & \textbf{0.96} & \textbf{0.96} & \textbf{1592} \\
\midrule

\multicolumn{6}{c}{\textbf{AdaBoost}} \\
\midrule
DDoS     & 1.00 & 0.92 & 0.95 & - & 872 \\
VoIP     & 1.00 & 1.00 & 1.00 & - & 221 \\
VideoTCP & 0.87 & 1.00 & 0.93 & - & 499 \\
\textbf{Overall} & \textbf{0.96} & \textbf{0.95} & \textbf{0.95} & \textbf{0.95} & \textbf{1592} \\
\midrule

\multicolumn{6}{c}{\textbf{CatBoost}} \\
\midrule
DDoS     & 0.98 & 0.94 & 0.96 & - & 872 \\
VoIP     & 1.00 & 1.00 & 1.00 & - & 221 \\
VideoTCP & 0.91 & 0.96 & 0.93 & - & 499 \\
\textbf{Overall} & \textbf{0.96} & \textbf{0.96} & \textbf{0.96} & \textbf{0.96} & \textbf{1592} \\
\midrule

\multicolumn{6}{c}{\textbf{Extra Trees}} \\
\midrule
DDoS     & 0.95 & 0.94 & 0.95 & - & 872 \\
VoIP     & 1.00 & 1.00 & 1.00 & - & 221 \\
VideoTCP & 0.90 & 0.92 & 0.91 & - & 499 \\
\textbf{Overall} & \textbf{0.94} & \textbf{0.94} & \textbf{0.94} & \textbf{0.94} & \textbf{1592} \\
\midrule

\multicolumn{6}{c}{\textbf{Random Forest}} \\
\midrule
DDoS     & 0.95 & 0.94 & 0.95 & - & 872 \\
VoIP     & 1.00 & 1.00 & 1.00 & - & 221 \\
VideoTCP & 0.90 & 0.92 & 0.91 & - & 499 \\
\textbf{Overall} & \textbf{0.94} & \textbf{0.94} & \textbf{0.94} & \textbf{0.94} & \textbf{1592} \\
\midrule

\multicolumn{6}{c}{\textbf{Gradient Boosting}} \\
\midrule
DDoS     & 0.96 & 0.94 & 0.95 & - & 872 \\
VoIP     & 1.00 & 1.00 & 1.00 & - & 221 \\
VideoTCP & 0.90 & 0.93 & 0.92 & - & 499 \\
\textbf{Overall} & \textbf{0.95} & \textbf{0.95} & \textbf{0.95} & \textbf{0.95} & \textbf{1592} \\
\midrule

\multicolumn{6}{c}{\textbf{Logistic Regression}} \\
\midrule
DDoS     & 0.85 & 0.84 & 0.84 & - & 872 \\
VoIP     & 1.00 & 1.00 & 1.00 & - & 221 \\
VideoTCP & 0.72 & 0.74 & 0.73 & - & 499 \\
\textbf{Overall} & \textbf{0.83} & \textbf{0.83} & \textbf{0.83} & \textbf{0.83} & \textbf{1592} \\
\midrule

\multicolumn{6}{c}{\textbf{Decision Tree}} \\
\midrule
DDoS     & 0.95 & 0.94 & 0.95 & - & 872 \\
VoIP     & 1.00 & 1.00 & 1.00 & - & 221 \\
VideoTCP & 0.90 & 0.92 & 0.91 & - & 499 \\
\textbf{Overall} & \textbf{0.94} & \textbf{0.94} & \textbf{0.94} & \textbf{0.94} & \textbf{1592} \\
\midrule

\multicolumn{6}{c}{\textbf{K-Nearest Neighbors}} \\
\midrule
DDoS     & 0.95 & 0.92 & 0.93 & - & 872 \\
VoIP     & 1.00 & 1.00 & 1.00 & - & 221 \\
VideoTCP & 0.87 & 0.91 & 0.89 & - & 499 \\
\textbf{Overall} & \textbf{0.93} & \textbf{0.93} & \textbf{0.93} & \textbf{0.93} & \textbf{1592} \\
\midrule

\multicolumn{6}{c}{\textbf{Support Vector Machine}} \\
\midrule
DDoS     & 0.84 & 0.82 & 0.83 & - & 872 \\
VoIP     & 1.00 & 1.00 & 1.00 & - & 221 \\
VideoTCP & 0.70 & 0.72 & 0.71 & - & 499 \\
\textbf{Overall} & \textbf{0.82} & \textbf{0.82} & \textbf{0.82} & \textbf{0.82} & \textbf{1592} \\
\midrule

\multicolumn{6}{c}{\textbf{Artificial Neural Network}} \\
\midrule
DDoS     & 1.00 & 0.91 & 0.95 & - & 872 \\
VoIP     & 1.00 & 1.00 & 1.00 & - & 221 \\
VideoTCP & 0.86 & 1.00 & 0.93 & - & 499 \\
\textbf{Overall} & \textbf{0.96} & \textbf{0.95} & \textbf{0.95} & \textbf{0.95} & \textbf{1592} \\

\end{longtable}

\subsection{Results Analysis}
This section analyzes the classification results obtained from various models, focusing on overall performance, robustness to class imbalance, and sources of misclassification. Key insights are drawn from evaluation metrics and confusion matrices to highlight model strengths and areas for improvement.

\subsubsection{Performance Interpretation}

The classification results obtained from the tested models are summarized in Table~\ref{tabref:table-3} and illustrated in Figure~\ref{fig:fig-8}. Overall, algorithms leveraging boosting methods (XGBoost, CatBoost, AdaBoost, and Gradient Boosting) along with ANN exhibited remarkable performance, achieving global F1-scores ranging between 0.95 and 0.96. Particularly, the XGBoost and CatBoost models demonstrated superior performance, each attaining an F1-score of 0.96, underscoring their effectiveness in capturing the characteristic patterns of network traffic within DDoS, VoIP, and VideoTCP scenarios.

Decision tree-based models (Random Forest, Extra Trees, and Decision Tree) also yielded robust results, each reaching an F1-score of 0.94. In contrast, simpler approaches such as Logistic Regression and Support Vector Machines exhibited relatively lower performances, achieving global F1-scores of 0.83 and 0.82, respectively. These findings confirm their limitations in effectively capturing complex interactions inherent to network traffic flows.

\subsubsection{Robustness to Class Imbalance}

The unequal distribution of data across the DDoS, VoIP, and VideoTCP classes represents a significant challenge to model robustness. Nevertheless, detailed results shown in Table~\ref{tabref:table-3} indicate that most tested models successfully maintained high precision and recall rates for the minority class (VoIP), often achieving scores close to 1.00. This outcome highlights the effectiveness of the adopted data rebalancing strategy via SMOTE, combined with the inherent robustness of the tested algorithms, in mitigating negative impacts caused by initial class imbalance. However, despite being less imbalanced than VoIP, the VideoTCP class consistently displayed slightly lower precision and recall scores. This observation suggests the persistent sensitivity of models to subtle intrinsic variations of VideoTCP flows, indicating a potential need for supplementary augmentation or specific data generation strategies targeting this particular class.

\subsubsection{Misclassification Analysis}

Confusion matrix analysis (cf. Table~\ref{tabref:table-3}) reveals that most classification errors predominantly occur between DDoS and VideoTCP classes, while the VoIP class is almost perfectly distinguished by all models. This finding indicates that flows associated with DDoS attacks and VideoTCP transmissions exhibit very similar characteristics regarding throughput and duration, complicating their clear differentiation by classification algorithms. Models such as XGBoost and CatBoost managed to significantly reduce these errors compared to other algorithms. However, the observed persistent confusion highlights the importance of better differentiating features used during training and suggests exploring hybrid approaches or advanced deep learning methods capable of capturing subtle distinctions between these two traffic types more effectively.

\section{Conclusion and Future Work}\label{sect:s6}

This paper presented a comprehensive evaluation of multiple machine learning techniques for detecting DDoS attacks in VANETs, specifically targeting emergency vehicle communication scenarios on highways. Leveraging a realistic simulation setup, which integrates the NS-3 network simulator with the SUMO mobility simulator and real-world vehicular mobility traces from Germany's A81 highway, we generated a robust and reproducible dataset for rigorous evaluation.

The experimental results demonstrated the high effectiveness of several machine learning algorithms, notably XGB, CB, AB, GB, and ANN, all achieving exceptional classification performance with F1-scores up to 96\%. Our findings confirmed the efficacy of SMOTE to handle imbalanced datasets, significantly enhancing the model’s ability to accurately classify minority classes, particularly the VoIP traffic class.

This study offers significant scientific contributions, including the introduction of a reproducible and realistic methodology combining NS-3 and SUMO simulators with authentic mobility data, and a systematic comparison of widely recognized machine learning classifiers in the context of highway VANET scenarios. Furthermore, the detailed SHAP-based feature selection analysis provided valuable insights into the key predictors necessary for accurate intrusion detection.

Despite these contributions, the study has several limitations. Primarily, the results remain constrained by the synthetic nature of the dataset, albeit enhanced by real-world mobility patterns. Moreover, the simulations did not encompass the full complexity of real-world communication scenarios, such as varying signal propagation conditions, diverse network topologies, and real-time network adaptations.

Future research should focus on extending the present approach through the following perspectives:
\begin{itemize}
    \item Conducting experiments in real-world settings by utilizing actual connected vehicles and infrastructure, which would validate and potentially refine the proposed classification models.
    \item Investigating the feasibility and effectiveness of deploying these detection systems onboard vehicles, thus enabling practical intrusion detection solutions in real-time scenarios.
    \item Expanding the methodology to detect other prominent cybersecurity threats in VANETs, including spoofing, Sybil, and blackhole attacks, thereby broadening the scope and practical applicability of the developed intrusion detection framework.
\end{itemize}

\section*{Acknowledgements}
None.

\section*{Funding}
The authors received no specific funding for this study.

\section*{Author Contributions}
Conceptualization, B.M.; Methodology, B.M. and V.F.; Software, B.M.; Investigation, B.M.; Writing—original draft, Bappa Muktar; Writing—review \& editing, V.F and N.A. All authors reviewed the results and approved the final version of the manuscript.

\section*{Availability of Data and Materials}
The data that support the findings of this study are available from the Corresponding Author, B.M., upon reasonable request.

\section*{Ethics Approval}
Not applicable.

\section*{Conflicts of Interest}
The authors declare no conflicts of interest to report regarding the present study.

\clearpage

\section*{Abbreviations}
\noindent The following abbreviations are used in this manuscript:\\

\begin{tabular}{@{}ll}
1D-CNN & One-Dimensional Convolutional Neural Network \\
AB     & AdaBoost \\
ANN    & Artificial Neural Network \\
CB     & CatBoost \\
DL     & Deep Learning \\
DDoS   & Distributed Denial of Service \\
DT     & Decision Tree \\
FDI    & False Data Injection \\
GB     & Gradient Boosting \\
GRU    & Gated Recurrent Unit \\
IDoS-CC & Intelligent DoS Attack Detection with Congestion Control \\
IoV    & Internet of Vehicles \\
KNN    & K-Nearest Neighbors \\
LR     & Logistic Regression \\
LSTM   & Long Short-Term Memory \\
ML     & Machine Learning \\
NS-3   & Network Simulator 3 \\
OMNeT++ & Objective Modular Network Testbed in C++ \\
OSM    & OpenStreetMap \\
RF     & Random Forest \\
RSU    & Roadside Unit \\
SD-VANET & Software-Defined Vehicular Ad Hoc Network \\
SDN    & Software Defined Networking \\
SHAP   & SHapley Additive exPlanations \\
SMOTE  & Synthetic Minority Over-sampling Technique \\
SNR    & Signal-to-Noise Ratio \\
SVM    & Support Vector Machine \\
SUMO   & Simulation of Urban MObility \\
TLBO   & Teaching and Learning-Based Optimization \\
UDP    & User Datagram Protocol \\
V2I    & Vehicle-to-Infrastructure \\
V2V    & Vehicle-to-Vehicle \\
VANET  & Vehicular Ad Hoc Network \\
VoIP   & Voice over IP \\
XGB    & XGBoost 
\end{tabular}

\bibliographystyle{unsrtnat}


\begin{thebibliography}{1}

	\bibitem{ref-1}
	Dutta, Arijit and Samaniego Campoverde, Luis Miguel and Tropea, Mauro and De Rango, Floriano. A comprehensive review of recent developments in vanet for traffic, safety \& remote monitoring applications. Journal of Network and Systems Management. 2024;32(4):73. [\href{https://doi.org/10.1007/s10922-024-09853-5}{CrossRef}]
	
	\bibitem{ref-2}
	Pawar, Vaishali and Zade, Nilima and Vora, Deepali and Khairnar, Vaishali and Oliveira, Aurenice and Kotecha, Ketan and Kulkarni, Ambarish. Intelligent Transportation System With 5G Vehicle-to-Everything (V2X): Architectures, Vehicular Use Cases, Emergency Vehicles, Current Challenges, and Future Directions. IEEE Access. 2024;12:183937--183960. [\href{https://doi.org/10.1109/ACCESS.2024.3506815}{CrossRef}]
	
	\bibitem{ref-3}
	Al-Mohtaseb, Abeer and Hanoon, Ali Qasim and Samara, Ghassan and Al Daoud, Essam and Alidmat, Omar and Batyha, Radwan and Aljaidi, Mohammad and Alazaidah, Raed and Elrashidi, Ali. A Comprehensive Review of VANET Attacks: Predictive Models, Vulnerability Management, and Defense Selection. 25th International Arab Conference on Information Technology (ACIT); 2024 Dec 10--12; Zarqa, Jordan. Piscataway, NJ, USA: IEEE; 2024. p. 1--9
	
	\bibitem{ref-4}
	Polat, Onur and Oyucu, Saadin and T{\"u}rko{\u{g}}lu, Muammer and Polat, H{\"u}seyin and Aksoz, Ahmet and Yard{\i}mc{\i}, Fahri. Hybrid AI-Powered Real-Time Distributed Denial of Service Detection and Traffic Monitoring for Software-Defined-Based Vehicular Ad Hoc Networks: A New Paradigm for Securing Intelligent Transportation Networks.  Applied Sciences. 2024;14(22):10501. [\href{https://doi.org/10.3390/app142210501}{CrossRef}]
	
	\bibitem{ref-5}
	Ababsa, Mohamed and Ribouh, Soheyb and Malki, Abdelhamid and Khoukhi, Lyes. Deep Multimodal Learning for Real-Time DDoS Attacks Detection in Internet of Vehicles.  arXiv preprint. 2025. [\href{https://doi.org/10.48550/arXiv.2501.15252}{CrossRef}]
	
	\bibitem{ref-6}
	Vamshi Krishna, K and Ganesh Reddy, K. Classification of distributed denial of service attacks in VANET: a survey. Wireless Personal Communications. 2023;132(2):933--964. [\href{https://doi.org/10.1007/s11277-023-10643-6}{CrossRef}]
	
	\bibitem{ref-7}
	Himanshu Setia, Amit Chhabra, Sunil K. Singh, Sudhakar Kumar, Sarita Sharma, Varsha Arya, Brij B. Gupta, Jinsong Wu. Securing the road ahead: Machine learning-driven DDoS attack detection in VANET cloud environments. Cyber Security and Applications. 2024;2:100037. [\href{https://doi.org/10.1016/j.csa.2024.100037}{CrossRef}]
	
	\bibitem{ref-8}
	Polat, Onur and Oyucu, Saadin and T{\"u}rko{\u{g}}lu, Muammer and Polat, H{\"u}seyin and Aksoz, Ahmet and Yard{\i}mc{\i}, Fahri. Hybrid AI-Powered Real-Time Distributed Denial of Service Detection and Traffic Monitoring for Software-Defined-Based Vehicular Ad Hoc Networks: A New Paradigm for Securing Intelligent Transportation Networks. Applied Sciences. 2024;14(22):10501. [\href{https://doi.org/10.3390/app142210501}{CrossRef}]
	
	\bibitem{ref-9}
	Polat, Huseyin and Turkoglu, Muammer and Polat, Onur. Deep network approach with stacked sparse autoencoders in detection of DDoS attacks on SDN-based VANET. IET Communications. 2020;14(22):4089--4100. [\href{https://doi.org/10.1049/iet-com.2020.0477}{CrossRef}]
	
	\bibitem{ref-10}
	Gopi, R and Mathapati, Mahantesh and Prasad, B and Ahmad, Sultan and Al-Wesabi, Fahd N and Alohali, Manal Abdullah and Hilal, Anwer Mustafa. Intelligent DoS Attack Detection with Congestion Control Technique for VANETs. Computers, Materials \& Continua. 2022;72(1):141--156. [\href{https://doi.org/10.32604/cmc.2022.023306}{CrossRef}]
	
	\bibitem{ref-11}
	Kadam, Nivedita and Krovi, Raja Sekhar. Machine Learning Approach of Hybrid KSVN Algorithm to Detect DDoS Attack in VANET. International Journal of Advanced Computer Science and Applications. 2021;12(7). [\href{https://dx.doi.org/10.14569/IJACSA.2021.0120782}{CrossRef}]
	
	\bibitem{ref-12}
	Alkadiri, Naam and Ilyas, Muhammad. Machine Learning-Based Architecture for DDoS Detection in VANETs System. 2022 International Conference on Artificial Intelligence of Things (ICAIoT); 2022 Dec 29--30;  Istanbul, Turkey. Piscataway, NJ, USA: IEEE; 2022. p. 1--7
	
	\bibitem{ref-13}
	Rashid, Kanwal and Saeed, Yousaf and Ali, Abid and Jamil, Faisal and Alkanhel, Reem and Muthanna, Ammar. An adaptive real-time malicious node detection framework using machine learning in vehicular ad-hoc networks (VANETs). Sensors. 2023;23(5):2594. [\href{https://doi.org/10.3390/s23052594}{CrossRef}]
	
	\bibitem{ref-14}
	Anyanwu, Goodness Oluchi and Nwakanma, Cosmas Ifeanyi and Lee, Jae-Min and Kim, Dong-Seong. Optimization of RBF-SVM kernel using grid search algorithm for DDoS attack detection in SDN-based VANET. IEEE Internet of Things Journal. 2022;10(10):8477--8490. [\href{https://doi.org/10.1109/JIOT.2022.3199712}{CrossRef}]
	
	\bibitem{ref-15}
	Marwah, Gagan Preet Kour and Jain, Anuj and Malik, Praveen Kumar and Singh, Manwinder and Tanwar, Sudeep and Safirescu, Calin Ovidiu and Mihaltan, Traian Candin and Sharma, Ravi and Alkhayyat, Ahmed. An improved machine learning model with hybrid technique in VANET for robust communication. mathematics. 2022;10(21):4030. [\href{https://doi.org/10.3390/math10214030}{CrossRef}]
	
	\bibitem{ref-16}
	Adhikary, Kaushik and Bhushan, Shashi and Kumar, Sunil and Dutta, Kamlesh. Hybrid algorithm to detect DDoS attacks in VANETs. Wireless Personal Communications. 2020;114(4):3613--3634. [\href{https://doi.org/10.1007/s11277-020-07549-y}{CrossRef}]
	
	\bibitem{ref-17}
	Tariq, Usman. Optimized Feature Selection for DDoS Attack Recognition and Mitigation in SD-VANETs. World Electric Vehicle Journal. 2024;15(9):395. [\href{https://doi.org/10.3390/wevj15090395}{CrossRef}]
	
	\bibitem{ref-18}
	Lekshmi V, R. Suji Pramila and Tibbie Pon Symon V A. Defense Mechanisms for Vehicular Networks: Deep Learning Approaches for Detecting DDoS Attacks. International Journal of Advanced Computer Science \& Applications. 2024;15(7). [\href{https://dx.doi.org/10.14569/IJACSA.2024.0150765}{CrossRef}]
	
	\bibitem{ref-19}
	Haydari, Ammar and Yilmaz, Yasin. RSU-based online intrusion detection and mitigation for VANET. Sensors. 2022;22(19):7612. [\href{https://doi.org/10.3390/s22197612}{CrossRef}]
	
	\bibitem{ref-20}
	Riley GF, Henderson TR. The ns-3 network simulator. In: Wehrle K, Güneş M, Gross J, editors. Modeling and tools for network simulation. Berlin, Heidelberg: Springer Berlin Heidelberg; 2010. p. 15--34.
	
	\bibitem{ref-21}
	Behrisch, Michael and Bieker, Laura and Erdmann, Jakob and Krajzewicz, Daniel. SUMO--simulation of urban mobility: an overview. In Proceedings of the SIMUL 2011, The Third International Conference on Advances in System Simulation; 2011 Oct 23--28; Barcelona, Spain. Red Hook, NY, USA: ThinkMind; 2011.
	
	\bibitem{ref-22}
	Chawla, Nitesh V and Bowyer, Kevin W and Hall, Lawrence O and Kegelmeyer, W Philip. SMOTE: synthetic minority over-sampling technique. Journal of artificial intelligence research. 2002;16:321--357. [\href{https://doi.org/10.1613/jair.953}{CrossRef}]

\end{thebibliography}





\end{document}